\documentclass[11pt,twoside]{article}
\usepackage{asp2010}

\resetcounters

\begin{document}

\title{Cone of Darkness: Finding Blank-sky Positions for Multi-object Wide-field Observations}
\author{Nuria~P.~F.~Lorente
\affil{Australian Astronomical Observatory, PO Box 915,\\
North Ryde, NSW 1670, Australia}
}

\begin{abstract}
We present the Cone of Darkness, an application to automatically configure blank-sky positions for a
series of stacked, wide-field observations, such as those carried out by the SAMI instrument on the
Anglo-Australian Telescope (AAT). The Sydney-AAO Multi-object Integral field spectrograph (SAMI)
uses a plug-plate to mount its $13 \times 61$ core imaging fibre bundles (hexabundles) in the optical plane
at the telescope's prime focus. To make the most efficient use of each plug-plate, several observing
fields are typically stacked to produce a single plate. When choosing blank-sky positions for the
observations it is most effective to select these such that one set of 26 holes gives valid sky
positions for all fields on the plate. However, when carried out manually this selection process is
tedious and includes a significant risk of error. The Cone of Darkness software aims to provide
uniform blank-sky position coverage over the field of observation, within the limits set by the
distribution of target positions and the chosen input catalogues. This will then facilitate the
production of the best representative median sky spectrum for use in sky subtraction. The
application, written in C++, is configurable, making it usable for a range of instruments. Given the
plate characteristics and the positions of target holes, the software segments the unallocated space
on the plate and determines the position which best fits the uniform distribution
requirement. This position is checked, for each field, against the selected catalogue using a TAP
ADQL search. The process is then repeated until the desired number of sky positions is
attained.
\end{abstract}

\section{Introduction}
The SAMI instrument \citep{2012clb+} has been developed to carry out a detailed galaxy study which
will use multi-object integral field spectroscopy techniques to study gas accretion in galaxies,
star formation as a function of environment, and the relationship between the star formation in a
galaxy and its AGN \citep{2012fbc+}.

SAMI uses 61-core hexabundles to observe a field of view of 15~arcsec for each of 13 targets
\citep{2012bbl+}. The hexabundles are held in place on the optical plane through the use of a plug
plate --- that is, a pre-drilled steel plate with holes corresponding to the positions of target
objects, for a given telescope pointing on the sky.
For observational efficiency and to maximise the utilisation of each plug plate, several (usually 2
or 3) consecutive fields to be observed are stacked and their positions of their targets are drilled
onto a single plate. This plate configuration process is carried out automatically, based on a pool
of targets for each field and on the known hardware constraints which govern the placement of holes
on the plate \citep{2013lfg}.

As well as the target positions, the plate configuration process allocates the instrument's 26 sky
fibres to blank sky positions in each field. To make the best use of the available space on the
plate, a sky plate hole is chosen such that it corresponds to blank sky for each of
the 2 (or more) stacked pointings on the plate. Additionally, the sky positions need to be as
uniformly distributed over the plate as the allocated target positions will allow, so as to
provide optimal sky subtraction data.

The Cone of Darkness (COD) application has been written
to carry out this task, in the first instance for SAMI, but with the flexibility and modularity to
determine blank sky positions for a general multi-object instrument.

\section{Cone of Darkness}
\begin{figure}[ht]
\centering
\includegraphics[width=\linewidth]{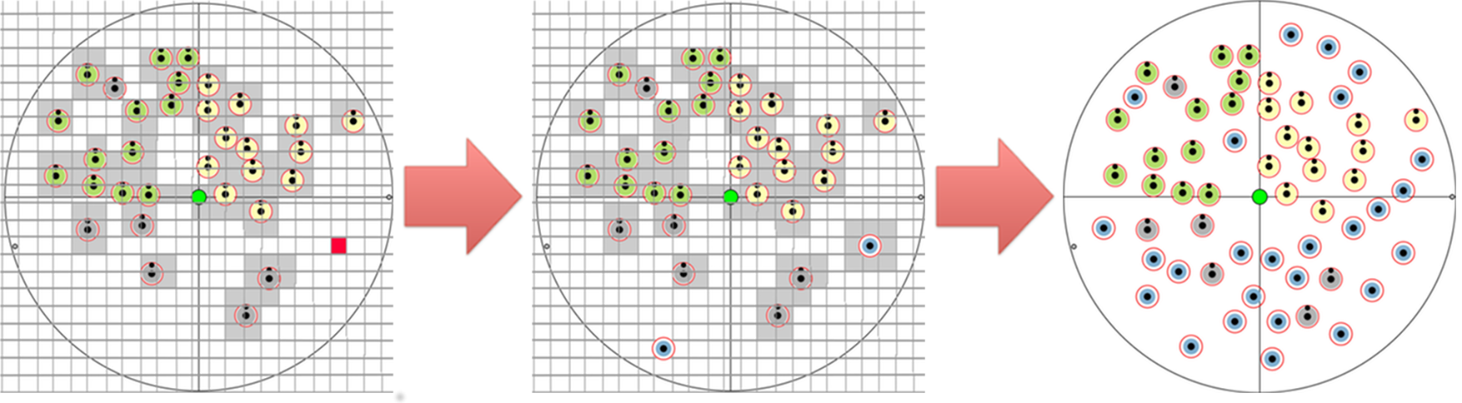}
\caption {Left: The plate area is split into cells and those containing a hole for targets in field
  1 (yellow), field 2 (green) or guide stars (grey) are excluded from the process (greyed-out
  cells). The most isolated remaining cell (red) becomes the sky position candidate.  Centre: The
  candidate position is checked for each stacked field. If no objects are found, the position is
  accepted (blue) and any cells containing the new hole are excluded from subsequent iterations.
  Right: The process continues until all requested sky fibres have been allocated a position.}
\label {O35_fig1}
\end{figure}

To produce optimal plate coverage of sky positions, a grid of user-defined size is laid over the
plate area, and the resulting grid cells form the set of candidate sky positions. From this set
those cells within the exclusion zone of any allocated position (i.e. the target holes plus an
instrument-specific area around them, to allow for connector size, etc.) are discarded. The most
remote candidate cell is then found, based on nearest-neighbour distances (Figure~\ref{O35_fig1},
left).

The candidate position is then tested for suitability, by checking whether there are any objects in the
field to a given magnitude limit, for each stacked field on the plate (2 fields in the
example in Figure~\ref{O35_fig1}).
This is done by means of a TAP ADQL search of the selected catalogue \citep{2011old+} which, in the
case of SAMI is the SuperCOSMOS Science Archive. Because there are 26 sky fibres to be allocated to
positions distributed over the entire plate, and because there are potentially many candidates to be
tested before each valid position is found, it is more efficient (due to network and query
overheads) to run a single catalogue search per field covering the entire one degree field-of-view
of the plate, rather than to execute a catalogue query for each individual candidate.\newpage
\noindent An example of such a TAP query is as follows:
{\footnotesize
\begin{verbatim}
http://wfaudata.roe.ac.uk/ssa-dsa/TAP/sync ? REQUEST=doQuery & LANG=ADQL &
QUERY=SELECT TOP 100 & QUERY=SELECT objID, ra, dec, epoch, sigRA, sigDec, 
muAcosD, muD, sigMuAcosD, sigMuD, chi2 FROM Source WHERE 
((ra BETWEEN (centralPosRADeg - radiusDeg / 2.0 AND 
(centralPosRADeg + radiusDeg / 2.0) AND 
(dec BETWEEN (centralPosDecDeg - radiusDeg / 2.0 AND  
(centralPosDecDeg + radiusDeg / 2.0)))
\end{verbatim}
}
The object list is returned by the query service as a VOTable \citep{2011owd+}, saved to disk and
then locally searched for objects within the field of view of each candidate sky position. This
currently takes the form of a radial proximity check, and in future versions will also include a
simple analysis of a 20~arcsec SuperCOSMOS thumbnail image of the region.

Once a candidate position is allocated to a sky fibre or discarded as unsuitable, the corresponding
grid cell is removed from the candidate pool. The nearest neighbour distances of remaining candidate
cells are recalculated and the most remote candidate is again identified. The process now iterates
(check candidate region for sources; acept or reject position; find next most isolated candidate
cell) until the required number of suitable sky positions are found.

To give the user a final way to confirm that the process has worked correctly --- i.e.\ that there
are no objects in the selected sky positions down to the detection limit of the instrument, the
program downloads a set of thumbnail images for the positions chosen. This allows a quick visual
inspection of the regions to be observed before the process of plate fabrication begins. 

\section{Future Work}
The Cone of Darkness application was written with the general fibre-positioning instrument in mind,
and the code is parametrised and abstracted to support other instruments with little extra
effor. However, its user interface is specific to SAMI and there are plans to make this more general
in subsequent versions of the code. Additionally, more flexible catalogue discovery, via a VO
Registry or through direct user input would be a very useful addition.  Finally, better
consideration of catalogue object morphology (circular objects are currently assumed) would result
in decreased probability of false positives, as well as better use of the available plate area.

\acknowledgements This work has made use of data obtained from the SuperCOSMOS Science Archive,
prepared and hosted by the Wide Field Astronomy Unit, Institute for Astronomy, University of
Edinburgh, which is funded by the UK Science and Technology Facilities Council.

\bibliography{O35}

\end{document}